\documentclass[twocolumn,amsmath,showpacs,amsfonts,aps,prc,floatfix]{revtex4}
\usepackage{graphicx}
\usepackage{bm}
\begin{document}
\title{An estimate of QGP viscosity from STAR data on $\phi$ mesons}
\author{A. K. Chaudhuri}
\email[E-mail:]{akc@veccal.ernet.in}
\affiliation{Variable Energy Cyclotron Centre, 1/AF, Bidhan Nagar, 
Kolkata 700~064, India}

\begin{abstract}
In the Israel-Stewart's theory of dissipative hydrodynamics,   
with a lattice based equation of state, where the confinement-deconfinement transition is a cross-over at $T_{co}$=196 MeV,  
we have analysed the STAR data on $\phi$ meson production in Au+Au collisions at $\sqrt{s}$=200 GeV. From a simultaneous fit to
$\phi$ mesons multiplicity, mean $p_T$ and integrated $v_2$, we obtain a phenomenological estimate of QGP viscosity, $\eta/s =0.15 \pm 0.05 \pm 0.03$,
the first error is due to the experimental uncertainty in STAR measurements, the second reflects the uncertainties in initial and final conditions of the fluid.
A host of STAR data, e.g. $\phi$ multiplicity, integrated $v_2$, mean $p_T$, $p_T$ spectra ($p_T <$3 GeV), in central Au+Au collisions, are consistent with the estimate of viscosity.
\end{abstract}


\date{\today}  

\maketitle

\section{Introduction} \label{sec1}
 
Experiments in Au+Au collisions at RHIC  \cite{BRAHMSwhitepaper,PHOBOSwhitepaper,PHENIXwhitepaper,STARwhitepaper}, 
produced convincing evidences that in non-central Au+Au collisions, a hot, dense, strongly interacting,  
collective QCD matter is created. Whether the matter can be characterized as the lattice QCD \cite{lattice,Cheng:2007jq} 
predicted Quark-Gluon-Plasma (QGP) or not,   is still a question of debate. A host of experimental data produced 
in Au+Au collisions at RHIC, at c.m. energy $\sqrt{s}$=200 GeV, have been successfully analysed using ideal 
hydrodynamics \cite{QGP3}. Multiplicity, mean $p_T$, $p_T$-spectra, elliptic flow etc.  of identified particles, 
are well explained in the ideal hydrodynamic model with QGP as the initial state.  Ideal hydrodynamics analysis 
of the RHIC data indicate that in central Au+Au collisions, at the equilibration time $\tau_i \approx$ 0.6 fm, 
central energy density of the QGP fluid is $\varepsilon_i \approx$30 $GeV/fm^{-3}$ \cite{QGP3}.

However, estimate of the initial condition of the fluid can not be creditable unless dissipative effects are accounted for. 
In recent years, considerable progress has been made in numerical implementation
 of dissipative hydrodynamics 
\cite{Teaney:2003kp,MR04,Koide:2007kw,Chaudhuri:2005ea,Heinz:2005bw,Chaudhuri:2006jd,Chaudhuri:2008je,Chaudhuri:2008sj,Chaudhuri:2007qp,Chaudhuri:2008ed,Romatschke:2007mq,Romatschke:2007jx,Song:2007ux,Song:2008si}. 
Unlike in ideal fluid evolution, where initial and final state entropy remains the same, in viscous fluid evolution entropy is generated. Consequently, to produce a fixed final state entropy, viscous fluid 
require less initial entropy density (or energy density) than an ideal fluid.
Amount of entropy generated depend on viscosity.   Viscosity of a strongly interacting QGP is quite uncertain. String theory based 
models (ADS/CFT) give a lower bound on viscosity of any matter $\eta/s \geq 1/4\pi$ \cite{Policastro:2001yc}. Perturbative QCD estimates $\eta/s \sim 1$ \cite{Arnold:2000dr}. At RHIC region, Nakamura and Sakai \cite{Nakamura:2005yf}
estimated the viscosity of a hot gluon gas  as $\eta/s$=0.1-0.4. 
In a SU(3) gauge theory, Meyer \cite{Meyer:2007ic} gave the upper bound $\eta/s <$1.0, and his best estimate 
is $\eta/s$=0.134(33) at $T=1.165T_c$. Given the uncertainty   ($1/4\pi \leq \eta/s \leq 1$), it is 
important to obtain a phenomenological estimate of viscosity of a strongly interacting QGP. Only then hydrodynamic analysis for the initial condition of the fluid will be reliable.

For long, strangeness enhancement is considered as a signature of QGP formation \cite {Koch:1986ud}. Compared to a hadron gas, in QGP, 
strangeness is enhanced due to abundant $gg\rightarrow s\bar{s}$ reactions. Early produced $s\bar{s}$, if survive
hadronisation can lead to increased production of strange particles compared to pp or pA collisions. Experimental data do show
strangeness enhancement \cite{strange}. 
Recently, STAR collaboration has measured $\phi$ meson production in Au+Au collisions at $\sqrt{s}$=200 GeV \cite{Abelev:2007rw}. Compared to pp collisions, in Au+Au collisions, $\phi$ meson production is enhanced. 
As noted in \cite{Mohanty:2009tz}, several unique features of $\phi$ mesons (e.g.
hidden strange particle, both hadronic and leptonic decay, not affected by resonance decays,   mass and width are not modified in a medium \cite{Alt:2008iv}  etc.) make it an ideal probe to investigate medium properties in heavy ion collisions. STAR data \cite{Abelev:2007rw} appear to be consistent with a model based on recombination of thermal strange quarks \cite{Hwa:2006vb}. 
As it will be shown below, STAR data on $\phi$ mesons are also consistent with hydrodynamic models and enable us to estimate QGP viscosity as $\eta/s=0.15 \pm 0.05 \pm 0.03$.

\section{Hydrodynamical Equations} \label{sec2}
In Israel-Stewart's theory of 2nd order dissipative hydrodynamics,  space-time evolution of a viscous  fluid is obtained by solving,  
 
\begin{eqnarray}  
\partial_\mu T^{\mu\nu} & = & 0,  \label{eq3} \\
D\pi^{\mu\nu} & = & -\frac{1}{\tau_\pi} (\pi^{\mu\nu}-2\eta \nabla^{<\mu} u^{\nu>}) \nonumber \\
&-&[u^\mu\pi^{\nu\lambda}+u^\nu\pi^{\nu\lambda}]Du_\lambda. \label{eq4}
\end{eqnarray}

Eq.\ref{eq3} is the conservation equation for the energy-momentum tensor, $T^{\mu\nu}=(\varepsilon+p)u^\mu u^\nu - pg^{\mu\nu}+\pi^{\mu\nu}$, 
$\varepsilon$, $p$ and $u$ being the energy density, pressure and fluid velocity respectively. $\pi^{\mu\nu}$ is the shear stress tensor (we have neglected bulk viscosity and heat conduction). Eq.\ref{eq4} is the relaxation equation for the shear stress tensor $\pi^{\mu\nu}$.   
In Eq.\ref{eq4}, $D=u^\mu \partial_\mu$ is the convective time derivative, $\nabla^{<\mu} u^{\nu>}= \frac{1}{2}(\nabla^\mu u^\nu + \nabla^\nu u^\mu)-\frac{1}{3}  
(\partial . u) (g^{\mu\nu}-u^\mu u^\nu)$ is a symmetric traceless tensor. $\eta$ is the shear viscosity and $\tau_\pi$ is the relaxation time.  It may be mentioned that in a conformally symmetric fluid relaxation equation can contain additional terms  \cite{Song:2008si}.

\begin{figure}[t]
 \center
 \resizebox{0.3\textwidth}{!}{%
  \includegraphics{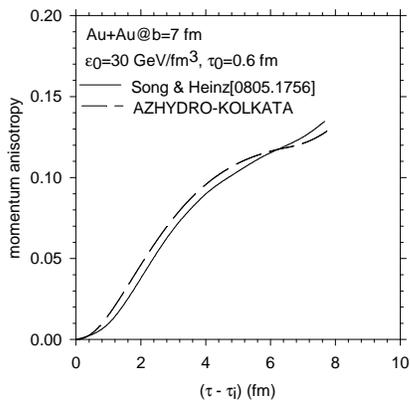}
}
\caption{Minimally viscous fluid ($\eta/s$=0.08) simulation for temporal evolution of momentum anisotropy in b=7 fm Au+Au collision at RHIC. The solid line is the simulation result from VISH2+1 \cite{Song:2008si} 
and the dashed line is the simulation result from AZHYDRO-KOLKATA.}  
\label{F1}
\end{figure}

Assuming boost-invariance, Eqs.\ref{eq3} and \ref{eq4}  are solved in $(\tau=\sqrt{t^2-z^2},x,y,\eta_s=\frac{1}{2}\ln\frac{t+z}{t-z})$ 
coordinates, with a code   "`AZHYDRO-KOLKATA"', developed at the Cyclotron Centre, Kolkata.
 Details of the code can be found in \cite{Chaudhuri:2008je,Chaudhuri:2008sj,Chaudhuri:2007qp}. In Fig.\ref{F1}, we have compared the temporal evolution 
 of   momentum anisotropy $\varepsilon_p=\frac{<T^{xx}-T^{yy}>}{<T^{xx}+T^{yy}>}$ of a QGP fluid with a calculation
 of Song and Heinz \cite{Song:2008si}. Initial conditions are similar for both the simulations.
Within 10\% or less, AZHYDRO-KOLKATA simulation  reproduces  Song and Heinz's  \cite{Song:2008si} result for temporal evolution of momentum anisotropy $\varepsilon_p$.

 \section{Equation of State}
\begin{figure}[t]
\center
 \resizebox{0.3\textwidth}{!}{%
  \includegraphics{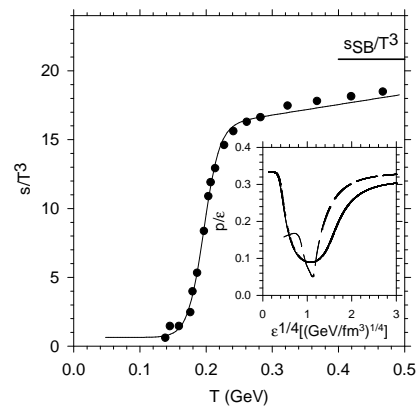}
}
\caption{Black circles are lattice simulation \cite{Cheng:2007jq} for entropy density. The black line is the parametric representation to the lattice simulations. In the inset, the solid and dashed lines are $p/\varepsilon$ in lattice based EOS and in an EOS incorporating 1st order transition \cite{QGP3}.}\label{F2}
\end{figure}

Equation of state (EOS) is one of the most important inputs of a hydrodynamic model. Through this input macroscopic hydrodynamic models make contact with the microscopic world. 
Most of the hydrodynamical calculations are performed with EOS with a 1st order phase transition.
Huovinen    \cite{Huovinen:2005gy} reported an 'ideal' hydrodynamic simulation with 2nd order phase transition. He concluded that the experimental data (e.g. elliptic flow of proton or antiproton) are better explained with EOS with 1st order phase transition than with EOS with 2nd order phase transition.
However, lattice simulations indicate that confinement to deconfinement transition is a cross-over at $\mu_B$=0 and in the $\mu_B$ range $\sim$30 MeV expected at RHIC \cite{Cheng:2007jq,Aoki:2006we}.  
In Fig.\ref{F2},  a recent lattice simulation  \cite{Cheng:2007jq} for the  entropy density is  shown.   The solid line in Fig.\ref{F2} is a parameterisation of the entropy density. 
  
\begin{equation}\label{eq1}
\frac{s}{T^3}=\alpha+[\beta+\gamma T][1+tanh\frac{T-T_{co}}{\Delta T}],
\end{equation}

From the parametric form of the entropy density, pressure and energy density can be obtained using the thermodynamic relations,

\begin{eqnarray}  
  p(T)&=&\int_0^T s(T) ds \label{eq2a} \\
  \varepsilon(T)&=&Ts -p \label{eq2b}.
  \end{eqnarray}
  
 Generally, in hydrodynamic simulations, hadronic phase is approximated by a (non-interacting) resonance hadron gas comprising all the resonances below 2-3 GeV. As the lattice simulation cover a wide temperature range below the cross-over temperature, $T_{co}=196(3)$ MeV,
we choose to use the   lattice based EOS (Eq.\ref{eq1}-\ref{eq2b}) both in the QGP and in the hadronic phase. 
 The confined phase is not a hadronic resonance gas. The  trace anomaly $(\varepsilon-3p)/T^4$ in the temperature range 140-200 MeV is approximately 30\% less than that of a hadronic resonance gas \cite{Cheng:2007jq}.  
In the inset of Fig.\ref{F2},  the ratio $p/\varepsilon$ in the lattice based EOS  is compared with $p/\varepsilon$ in an   EOS
with 1st order phase transition \cite{QGP3}, which model the quark phase with bag model, and the hadronic phase by the hadronic resonance gas. In 1st order EOS, $p/\varepsilon$ fall sharply near the critical temperature. The fall is smoothened out in cross-over  transition. Lattice based EOS is also softer
at high temperature but harder at low temperature as compared to a first order phase transition.

\section{Initial conditions}

Boost-invariant solution of ideal hydrodynamics require transverse profile of the energy density ($\varepsilon(x,y)$) and
fluid four velocity ($u(x,y)$) at the initial time ($\tau_i$). A freeze-out prescription, e.g. freeze-out temperature $T_F$
is also needed.  In viscous hydrodynamics, at the initial time ($\tau_i$), additionally required are 
(independent) shear stress tensor ($\pi^{\mu\nu}(x,y)$) components. One also needs to specify 
viscosity coefficient $\eta$ and  the relaxation time $\tau_\pi$.
In the present analysis,
initial time for the hydrodynamic simulation is chosen to be $\tau_i$=0.2 fm.
Hydrodynamics analysis of photon data in RHIC Au+Au collisions suggest small initial time, $\tau_i$=0.2 fm 
\cite{Chatterjee:2008tp}. Even smaller formation time is suggested in the analysis of $J/\psi$ suppression at RHIC \cite{Chaudhuri:2007qz}. Fluid is assumed to have zero velocity $u(x,y)=0$ at the initial time. 
The initial energy density is assumed to be distributed as \cite{QGP3},

\begin{figure}[t]
 \center
 \resizebox{0.3\textwidth}{!}{%
  \includegraphics{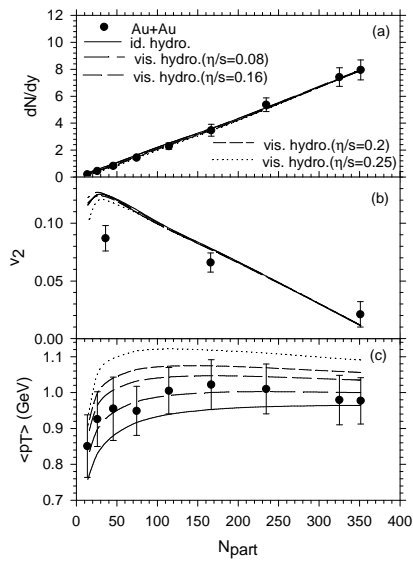}
}
\caption{STAR data on the centrality dependence of $\phi$ meson (a) multiplicity, (b) integrated $v_2$ and (c) mean $p_T$
are compared with hydrodynamical simulation of ideal and viscous fluid.}  
\label{F3}
\end{figure}

\begin{equation} \label{eq6}
\varepsilon({\bf b},x,y)=\varepsilon_i[0.75 N_{part}({\bf b},x,y) +0.25 N_{coll}({\bf b},x,y)],
\end{equation}

\noindent
where ${\bf b}$ is the impact parameter of the collision and $N_{part}$ and $N_{coll}$ are the   average participant and collision number respectively.  The constant $\varepsilon_i$ does not depend on collision centrality. Effect of viscosity is enhanced with non-zero initial shear stress
tensor \cite{Chaudhuri:2008sj} and we initialise it to boost-invariant value. The independent components are initialsied as, $\pi^{xx}=\pi^{yy}=2\eta/3\tau_i$, $\pi^{xy}=0$. For the relaxation time $\tau_\pi$, we use the Boltzmann
approximation $\tau_\pi=3\eta/2p$. Finally, in the cooper-Frye prescription, invariant yield for the $\phi$ mesons is calculated at the freeze-out temperature, $T_F$=140 MeV  
\cite{note1}.

 \begin{table}[ht]
\caption{\label{table1} The fitted values of the initial central energy density ($\varepsilon_i$) and temperature ($T_i$) of the fluid in b=0 Au+Au collisions, for different values of viscosity to entropy ratio ($\eta/s$). In the last three rows, $\chi^2/N$ for STAR data on $\frac{dN^\phi}{dy}$, $v_2$ and $<p_T^\phi>$ are shown.} 
\begin{ruledtabular} 
  \begin{tabular}{|c|c|c|c|c|c|}\hline
$\eta/s$         & 0    & 0.08 & 0.16 & 0.20 & 0.25 \\  \hline
$\varepsilon_i (GeV/fm^3)$ & $32.9$ & $20.0$ & $11.6$ &  $8.8$ &$6.15$ \\  
  & $\pm$ 3.7 & $\pm$ 2.4 & $\pm$ 1.4 &  $\pm$ 1.2 & $\pm$0.85 \\ \hline
$T_i$ (MeV) & 369.3 & 326.4 & 284.2 & 264.5  & 241.6   \\ 
  & $\pm 10.2$ & $\pm 9.8$ & $\pm 8.9$ & $\pm 9.4$  & $\pm 8.1$   \\ \hline  
$\chi^2/N (dN/dy)$ & 2.9 & 2.1 &  0.19 & 0.32 & 1.89\\ \hline
$\chi^2/N (<p_T>)$ & 0.71 & 0.01 &  0.49 & 1.14 & 2.79\\ \hline
$\chi^2/N (v_2)$ & 4.5 & 4.9 &  4.8 & 4.4 & 3.7 \\ \hline
\end{tabular}\end{ruledtabular}  
\end{table}

\section{Results}\label{sec3}

 Central energy density ($\varepsilon_0$) and fluid viscosity ($\eta$) are the parameters in our analysis. We assume that viscosity to entropy ratio $\eta/s$ remain constant throughout the evolution and simulate Au+Au collisions for   $\eta/s$=0, 0.08, 0.16, 0.2 and 0.25. 
We fix the central energy density of the fluid to reproduce the $\phi$ meson multiplicity in 0-5\% centrality Au+Au collisions. The corresponding central energy density and temperature  are listed in table.\ref{table1}. The uncertainty in $\varepsilon_i$ is due to the statistical and systematic error in STAR measurements. Due to entropy generation in viscous dynamics,  
initial energy density or temperature  is less in viscous fluid than in ideal fluid.  

In Fig.\ref{F3}, in three panels,
STAR data \cite{Abelev:2007rw} on the centrality dependence of $\phi$ meson (a) multiplicity ($\frac{dN^\phi}{dy}$), (b) integrated $v_2$ and (c) mean $p_T$ ($<p_T^\phi>$) are shown.  
Ideal or viscous fluid, initialised to fit $\phi$ meson multiplicity in 0-5\% collisions, reproduces the STAR data on $\phi$ multiplicity in
all the centrality ranges of collisions. STAR collaboration measured integrated $v_2$ only in 0-5\%, 10-40\% and 40-80\% centrality collisions. 
Fluid dynamics overestimate the flow in very peripheral collisions. In central or mid central collisions, the integrated $v_2$ is reasonably well reproduced both in ideal and viscous fluid evolution. From Fig.\ref{F3}c, we also observe that  the STAR data on $<p_T^\phi>$ are not explained unless $\eta/s <$0.25, otherwise it is overestimated. $\chi^2/N$ of the combined data sets (see table.\ref{table1}), enable us to estimate QGP viscosity as $\eta/s$=0.15 $\pm$ 0.05.
 
The estimate does depend on the assumed initial and final conditions, e.g. initial time, initial velocity, freeze-out temperature etc.
Other conditions remaining unchanged, in viscous fluid ($\eta/s$=0.16) evolution,
$<p_T^\phi>$ decreases by $\sim$ 10\% if initial time increases from 
$\tau_i$=0.2 fm to 1.0 fm. Simultaneously, $\frac{dN^\phi}{dy}$ increases. Empirically, $<p_T^\phi>$ linearly increases with $\eta/s$.  
Then for higher initial time, initial temperature of the fluid will be reduced, but $\eta/s$ will increase by $\sim$10\%. $\eta/s$ will decrease by 2-4\% if initially fluid has small transverse velocity, $v_r=tanh(\alpha r)$, $\alpha$=0.02-0.04.
With small initial transverse velocity  $<p_T^\phi>$ is increased, multiplicity remaining unchanged. 
$<p_T^\phi>$ increase (decrease) with decreasing (increasing) $T_F$. $\frac{dN^\phi}{dy}$ shows an opposite trend. If $T_F$ ranges between 130-150 MeV,   $\eta/s$  will be uncertain by $\sim$8\%.  
In AZHYDRO-KOLKATA, fluid evolution is computed with $\sim$5\% accuracy leading to $\sim$9\% uncertainty in $\eta/s$. They all add to the systematic error, and we estimate the viscosity to entropy ratio of QGP fluid as, $\eta/s=0.15 \pm 0.05 \pm 0.03$. 
The estimate will not be affected if hard collision contribution in initial energy density increases. In 0-5\% Au+Au collisions, increasing hard collision contribution to initial energy density from $0 \rightarrow$ 75\%,
though multiplicity is reduced, $<p_T^\phi>$ remain essentially unchanged. More hard collision contribution will increase initial fluid temperature.  

\begin{figure}[t]
\center
 \resizebox{0.3\textwidth}{!}{
  \includegraphics{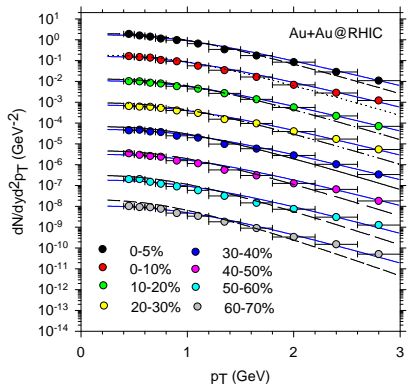}
}
\caption{STAR data \cite{Abelev:2007rw} on $p_T$ spectra of $\phi$ mesons in different centrality ranges of Au+Au collisions. The dashed black and solid blue lines are $\phi$ spectra from evolution of ideal and viscous ($\eta/s$=0.15) fluid respectively.}
\label{F4}
\end{figure}

As shown in Fig.\ref{F4},   $\phi$ mesons $p_T$ spectra in central and mid central Au+Au collisions are better reproduced in viscous ($\eta/s$=0.15) fluid evolution than in ideal fluid evolution. We have not shown here,   but 
$\eta/s$=0.15 is also consistent with experimental data on pion, kaon etc.

\section{Summary and Conclusions} \label{sec4} 

To summaries,  from the STAR data on the centrality dependence    of $\phi$ mesons multiplicity, mean $p_T$ and integrated $v_2$, we have obtained  a phenomenological estimate of QGP viscosity, $\eta/s = 0.15 \pm 0.05 \pm 0.03$, the   first error corresponds to experimental uncertainty in STAR measurements, the 2nd   due to the uncertainty in initial and final condition of the fluid. We conclude that low $p_T \leq$ 3 GeV, $\phi$ meson data in central Au+Au collisions are consistent with hydrodynamical evolution of QGP fluid with viscosity to entropy ratio  $\eta/s \approx $ 0.15.


\begin{thebibliography}{99}
\bibitem{BRAHMSwhitepaper}
 BRAHMS Collaboration, I. Arsene {\it et al.},  
Nucl. Phys. A {\bf 757}, 1 (2005). 
 
\bibitem{PHOBOSwhitepaper} 
PHOBOS Collaboration,  B. B. Back {\it et al.},  
Nucl. Phys. A {\bf 757}, 28 (2005). 
 
\bibitem{PHENIXwhitepaper} 
PHENIX Collaboration, K.~Adcox {\it et al.}, 
Nucl. Phys. A {\bf 757} 184 (2005).  
  
\bibitem{STARwhitepaper} 
STAR Collaboration, J. Adams {\it et al.}, 
Nucl. Phys. A {\bf 757} 102 (2005).

\bibitem{lattice} 
Karsch F, Laermann E, Petreczky P, Stickan S and Wetzorke I, 
2001 {\it Proccedings of NIC Symposium} (Ed. H. Rollnik and D. Wolf, John 
von Neumann Institute for Computing, J\"ulich, NIC Series, vol.9, 
ISBN 3-00-009055-X, pp.173-82,2002.)

\bibitem{Cheng:2007jq}
  M.~Cheng {\it et al.},
  Phys.\ Rev.\  D {\bf 77}, 014511 (2008).
  
\bibitem{QGP3}
P.~F. Kolb and U. Heinz,
in {\it Quark-Gluon Plasma 3}, edited by R.~C. Hwa and 
X.-N. Wang (World Scientific, Singapore, 2004), p.~634.

  

  
\bibitem{Teaney:2003kp}
  D.~Teaney,
  Phys.\ Rev.\  C {\bf 68}, 034913 (2003).

\bibitem{MR04}  A.~Muronga and D.~H.~Rischke,
  nucl-th/0407114\,(v2).
\bibitem{Koide:2007kw}
  T.~Koide, G.~S.~Denicol, Ph.~Mota and T.~Kodama,
  Phys.\ Rev.\  C {\bf 75}, 034909 (2007).
  
\bibitem{Chaudhuri:2005ea}
  A.~K.~Chaudhuri and U.~W.~Heinz,
  J.\ Phys.\ Conf.\ Ser.\  {\bf 50}, 251 (2006).


\bibitem{Heinz:2005bw}
  U.~W.~Heinz, H.~Song and A.~K.~Chaudhuri,
  Phys.\ Rev.\  C {\bf 73}, 034904 (2006).
  
\bibitem{Chaudhuri:2006jd}
  A.~K.~Chaudhuri,
  Phys.\ Rev.\  C {\bf 74}, 044904 (2006).
\bibitem{Chaudhuri:2008je}
 A.~K.~Chaudhuri,
  Phys. Lett. B672, 126, 2009.
\bibitem{Chaudhuri:2008sj} A.~K.~Chaudhuri,
 arXiv:0801.3180 [nucl-th].
\bibitem{Chaudhuri:2007qp} A.~K.~Chaudhuri,
  arXiv:0708.1252 [nucl-th].
  \bibitem{Chaudhuri:2008ed} A.~K.~Chaudhuri,
  J.\ Phys.\ G {\bf 35}, 104015 (2008).

\bibitem{Romatschke:2007mq}
  P.~Romatschke and U.~Romatschke,
  Phys.\ Rev.\ Lett.\  {\bf 99}, 172301 (2007).
\bibitem{Romatschke:2007jx}
  P.~Romatschke,
  Eur.\ Phys.\ J.\  C {\bf 52}, 203 (2007).

  
\bibitem{Song:2007ux}
  H.~Song and U.~W.~Heinz,
  Phys.\ Rev.\  C {\bf 77}, 064901 (2008), Phys.\ Rev.\  C {\bf 78}, 024902 (2008)..
  
\bibitem{Song:2008si}
  H.~Song and U.~W.~Heinz,
  Phys.\ Rev.\  C {\bf 78}, 024902 (2008)
  [arXiv:0805.1756 [nucl-th]].
  
\bibitem{Policastro:2001yc}
  G.~Policastro, D.~T.~Son and A.~O.~Starinets,
  Phys.\ Rev.\ Lett.\  {\bf 87}, 081601 (2001).
  
\bibitem{Arnold:2000dr}
  P.~Arnold, G.~D.~Moore and L.~G.~Yaffe,
  JHEP {\bf 0011}, 001 (2000),JHEP {\bf 0305}, 051 (2003).
  
 
  
\bibitem{Nakamura:2005yf}
  A.~Nakamura and S.~Sakai,
  Nucl.\ Phys.\  A {\bf 774}, 775 (2006).
\bibitem{Meyer:2007ic}
  H.~B.~Meyer,
  Phys.\ Rev.\  D {\bf 76}, 101701 (2007).
  
  
\bibitem{Koch:1986ud}
  P.~Koch, B.~Muller and J.~Rafelski,
  Phys.\ Rept.\  {\bf 142}, 167 (1986).
  
  \bibitem{strange}
  Chen J. H.  (for the STAR collaboration) J. Phys. G: Nucl. Part. Phys. 35, 104053 (2008).
  

  
\bibitem{Abelev:2007rw}
  B.~I.~Abelev {\it et al.}  [STAR Collaboration],
  Phys.\ Rev.\ Lett.\  {\bf 99}, 112301 (2007),arXiv:0809.4737 [nucl-ex],arXiv:0810.4979 [nucl-ex].
    
\bibitem{Mohanty:2009tz}
  B.~Mohanty and N.~Xu,
  arXiv:0901.0313 [nucl-ex].
\bibitem{Alt:2008iv}
  C.~Alt {\it et al.}  [NA49 collaboration],
  Phys.\ Rev.\  C {\bf 78}, 044907 (2008)
  [arXiv:0806.1937 [nucl-ex]].
  
\bibitem{Hwa:2006vb}
  R.~C.~Hwa and C.~B.~Yang,
  arXiv:nucl-th/0602024.


\bibitem{Huovinen:2005gy}
  P.~Huovinen,
  Nucl.\ Phys.\  A {\bf 761}, 296 (2005).
\bibitem{Aoki:2006we}
  Y.~Aoki, G.~Endrodi, Z.~Fodor, S.~D.~Katz and K.~K.~Szabo,
  Nature {\bf 443}, 675 (2006)
  [arXiv:hep-lat/0611014].
\bibitem{Chatterjee:2008tp}
  R.~Chatterjee and D.~K.~Srivastava,
  arXiv:0809.0548 [nucl-th].
\bibitem{Chaudhuri:2007qz}
  A.~K.~Chaudhuri,
  Phys.\ Lett.\  B {\bf 655}, 241 (2007).
  
\bibitem{note1} We have checked that with the lattice based EOS, in ideal fluid dynamics, STAR measurements of $\frac{dN^\phi}{dy}$ and $<p_T^\phi>$ in 0-5\% Au+Au collisions are best explained only with $T_F$=140 MeV.  Higher or lower $T_F$ fails to  reproduce $\frac{dN^\phi}{dy}$ and $<p_T^\phi>$ simultaneously.     
\end{thebibliography}
\end{document}